\documentstyle{article}
\begin{document}
\title{
Quantum electrodynamics in the squeezed vacuum state:
Feynman rules and corrections to the electron mass
 and anomalous magnetic moment
}
\author{K. Svozil\\
 {\small Institut f\"ur Theoretische Physik}  \\
  {\small Technische Universit\"at Wien   }     \\
  {\small Wiedner Hauptstra\ss e 8-10/136}    \\
  {\small A-1040 Vienna, Austria   }            \\
  {\small e1360dab@awiuni11.edvz.univie.ac.at}}
\maketitle
\begin{abstract}
Due to the nonvanishing average photon population of the squeezed vacuum
state, finite corrections to the scattering matrix are obtained.  The
lowest order contribution to the electron mass shift for a one mode
squeezed vacuum state is given by $\delta m(\Omega ,s)/m=\alpha (2/\pi
)(\Omega /m)^2\sinh^2(s)$, where $\Omega$ and $s$ stand for the mode
frequency and the squeeze parameter and $\alpha$ for the fine structure
constant, respectively.  The correction to the anomalous magnetic moment
of the electron is $\delta a_e(s)=-(4\alpha /\pi )\sinh^2(s)$.
 \end{abstract}

The dependece of the scattering matrix on the vacuum state
of the theory and on exterior parameters has been studied for the
 thermal
equilibrium \cite{thermal}, in cavity--quantum
electrodynamics \cite{cavity-qed},
on fractal space--time support \cite{zei-svo}
and, to some extent, in the presence of strong electromagnetic
fields \cite{greiner,strongfield-qed}.
Here, quantum electrodynamics
is investigated
in the presence of squeezed vacuum fluctuations \cite{milburn}; i.e.,
 fluctuations
with reduced noise in amplitude or phase.

The squeezed vacuum state \cite{loudon} exhibits
a nonvanishing average photon density proportional to $\sinh^2(s)$ per
squeezed mode, where $s$ is the squeeze parameter \cite{fn1}.
This can be accounted for in the perturbation series by the introduction
of a causal photon propagator as follows \cite{f1}.
 Denote the squeezed vacuum by $\vert {\rm sv}\rangle $. The
 photon propagator in the Landau gauge is
 \begin{eqnarray}
 D_{\mu \nu}(x-y)&=&-i\langle {\rm sv}\vert T[A_\mu (x)A_\nu (y)]\vert
 {\rm sv}\rangle \nonumber \\
 &=&ig_{\mu \nu}\int {d^3k\over (2\pi )^3}{{dk'}^3\over
 2(E_kE_{k'})^{1/2}}\langle {\rm sv}\vert \theta (x_0-y_0)[e^{-i(k
 x -k' y)}a_ka^\dagger_{k'}+e^{i(k x-k'
 y)}a^\dagger_ka_{k'}]+ x\leftrightarrow y \vert {\rm sv}\rangle
 \nonumber \\
 &=&ig_{\mu \nu}\left\{ \int {d^3k\over (2\pi )^3}{1\over 2E_k}[\theta
 (x_0-y_0)e^{-ik (x-y)}+\theta (y_0-x_0)e^{ik (x-y)}]+
 \right.
 \nonumber \\
 & &\qquad \qquad +  \left.\int
 {d^3k\over (2\pi )^3}{1\over 2E_k}
 n(k)[e^{ik (x-y)}+e^{-ik (x-y)}]\right\}
 ,
 \end{eqnarray}
 where the $aa^\dagger$ terms generate the usual causal propagator
 while the
 $a^\dagger a$ terms count the particle density in the squeezed vacuum.
 The propagator can be rewritten
  using contour--integral techniques
\begin{eqnarray}
D_{\mu \nu}(x-y)&=& \int {d^4k\over (2\pi )^4}e^{-ik(x-y)}D_{\mu \nu }
 (k)\nonumber \\
D_{\mu \nu }(k)&=&-g_{\mu \nu }\left[{1\over k^2+i\epsilon }-2\pi i
\delta (k^2)n(k)\right] \quad .                     \label{prop}
\end{eqnarray}
For the one mode squeezed state, $n(k;\Omega ,s)=\Omega \sinh^2(s)
\delta (
E_k -\Omega )$, where $E_k$ is the photon energy parameter and $\Omega$
 and $s$ stand for the frequency of
the squeezed mode and the squeezing parameter, respectively.
The  electron propagator $S(p)=1/(\rlap{/}p -m+i\epsilon )$, as well
as the bare vertex $\gamma_\mu$ remain unchanged.
 Notice however that a preferred frame of
 reference has been introduced due to the noncovariant choice of the
 density $n(k;\Omega ,s)$, i.e., the one at rest with respect to
 the squeezed vacuum. The resulting breakdown of Lorentz invariance
 necessitates a careful interpretation of the usual renormalisation
 prescriptions.

In what follows, the lowest order correction to the radiative mass of
the electron will be calculated. This can be done by evaluating
the second order contribution to the self energy
of the electron
\begin{equation}
 \Sigma (p;\Omega ,s)=-ie^2\int{d^4k\over (2\pi )^4}[iD_{\mu
\nu}(k;\Omega ,s)]\gamma^\mu {i\over \rlap{/}p-\rlap{/}k-m}\gamma^\nu
 \quad
 .
\end{equation}
The physical mass is interpreted as usual as the pole of the
renormalized electron propagator. For $\delta m(\Omega ,s)\ll m$,
\begin{eqnarray}
m(\Omega ,s)&\approx &m+\delta m+\Sigma (p;\Omega
 ,s)\vert_{\rlap{/}p=m}\nonumber \\
 & &\qquad =m+\delta m+\Sigma
(p;s=0)\vert_{\rlap{/}p=m}+\delta \Sigma
(p;\Omega,s)\vert_{\rlap{/}p=m}\nonumber \\
 & &\qquad =m+\delta m(\Omega ,s) \quad ,
\end{eqnarray}
where $m$ stands for the renormalized unsqueezed mass of the electron.

The correction term $\delta m(\Omega ,s) =\delta
\Sigma(p;\Omega ,s)\vert_{\rlap{/}p=m}$
due to squeezing adds up coherently to the renormalization
contributions of $m$. Its explicit form is given by
\begin{eqnarray}
\delta m(\Omega ,s)&=&-{e^2\over (2\pi )^3}\int d^4k \delta
(k^2)n(k;\Omega ,s)\gamma_\mu
{\rlap{/}p-\rlap{/}k+m\over  (p-k)^2-m^2+i\epsilon }\gamma^\mu
 \mid_{\rlap{/}p=m}         \nonumber  \\
 &=&{\alpha \over 2\pi^2}{
 I_\mu (p)p^\mu \over m}\vert_{p^2=m^2}\quad ,
\end{eqnarray}
 where
 Gordon's identity which reduces to $\gamma_\mu =p_\mu /m$
 has been used,
 $\alpha =e^2/4\pi$ stands for the fine structure constant and
\begin{equation}
 I_\mu (p)=\int {d^3\vec {\bf k}}{k_\mu
 \over  \vert \vec  {\bf k}\vert (pk)}n(\vert \vec {\bf k}\vert ;\Omega
,s)
\quad .
 \end{equation}
 In the rest frame of the electron
this expression can be  evaluated, yielding
\begin{equation}
{\delta m(\Omega ,s)/ m}= \alpha (2 / \pi)
(\Omega / m)^2\sinh^2(s)\quad .
\end{equation}
For optical frequencies, $\delta m(s)/m\approx 10^{-13}\sinh^2(s)$.

 The correction $\delta a_e$ to the anomalous
 magnetic
 moment of the electron $a_e$ can be extracted from a decomposition of
 the vertex function
 $\Lambda_\mu =\gamma_\mu+\Gamma_\mu$ on shell
 \begin{equation}
 \bar u(p-q)\Lambda_\mu u(p)=\bar u(p-q)\left[ \gamma_\mu
 f_1(q^2)+{i\over 2m}\sigma_{\mu \nu}q^\nu f_2(q^2)\right] u(p)
 \end{equation}
 with $a_e=f_2(q^2=0)$. To lowest order one obtains
 \begin{eqnarray}
 \Gamma_\mu (p,p-q;\Omega ,s)
 &=&
 \Gamma_\mu (p,p-q;s=0)+\delta
 \Gamma_\mu (p,p-q;\Omega ,s)  \nonumber \\
 &=&(-ie)^2\int {d^4k\over (2\pi
 )^4}[iD_{\alpha \beta }(k;\Omega ,s)]\gamma^\alpha {i\over
 \rlap{/}p-\rlap{/}q-\rlap{/}k-m}\gamma_\mu {i\over
 \rlap{/}p-\rlap{/}k-m}\gamma^\beta  .
 \end{eqnarray}
 The correction due to squeezing for $p=(m,0,0,0)$ and
 $q=(q_0,0,0,0)$, $q_0^2\ll m^2$ can be written as
 \begin{eqnarray}
  \delta
 \Gamma_\mu (p,p;\Omega ,s)
 &=&
 -(i\alpha /2\pi^2)(m^2I\gamma_\mu -I_{\mu \nu}\gamma^\nu )
    \label{gamma}
 \\
 I&=&
 \int d^4k {\delta (k^2)n(k;\Omega ,s)
 \over (pk)^2}
 \\
 I_{\mu \nu }&=&
 \int d^4k {k_\mu k_\nu \delta (k^2)n(k;\Omega ,s)
 \over (pk)^2}
 \quad .
 \end{eqnarray}
 For a definition
 of $a_e$ the term proportional to $\sigma_{\mu \nu}q^\nu$ is chosen.
 Hence, only the first
 term proportional to $I\gamma_\mu$ on the right hand side of
 (\ref{gamma}) is relevant.
 Using Gordon's identity, which is $\gamma_\mu =(1/2m)(2p_\mu
 -q_\mu -i\sigma_{\mu \nu}q^\nu)$ here, one obtains for the
 correction to the anomalous magnetic moment of the electron
 \begin{equation}
 \delta a_e(s)=-(4\alpha /\pi )\sinh^2(s) .
 \end{equation}

 This correction to the anomalous magnetic moment of the electron
 becomes comparable
 to the unsqueezed value $a_e=\alpha /2\pi $ for $s\approx 0.35$
 and increases rapidly as the population of the
 squeezed vacuum increases and should be
 testable with realistic experimental parameters.


\begin{thebibliography}{99}
\bibitem{fn1}
 other squeezing parameters associated with more general canonical
 transformations [isomorphic to elements of the real symplectic
 group $Sp(2N,R)$ for $N$--mode squeezing]
 have been set zero, but a generalisation is
 straightforward.
\end{thebibliography}
\end{document}